\documentclass[prd,aps,showpacs,secnumarabic]{revtex4}
\usepackage[dvips]{graphicx,color}
\begin{document}
\def\ds{\displaystyle}
\def\ss{\scriptstyle}
\def\sb{\mbox{\rule{0pt}{8pt}}}
\def\hh{\mbox{\rule[-6pt]{0pt}{20pt}}}
\title{Constraints on the dark energy equation of state from recent supernova data }
\author{Duane A. Dicus} \affiliation{Center for Particle Physics, University of Texas, Austin,
TX 78712}\author{Wayne W. Repko} \affiliation{Department of Physics and
Astronomy, Michigan State University, East Lansing, MI 48824}

\date{\today}

\begin{abstract}
   Measurements suggest that our universe has a substantial dark energy component.  The most recent data on type Ia supernovae give a dark energy density which is in good agreement with other measurements if the dark energy is assumed to be a cosmological constant. Here we examine to what extent that data can put constraints on a more general equation of state for the dark energy.

\end{abstract}
\maketitle

\section{Introduction}

It is generally believed, based on the analysis of type Ia supernovae red shift data \cite{hiz,super}, that the expansion of the universe is accelerating and that this acceleration is either caused by a ``dark'' energy or is the effect of extra dimensions.  In the case of dark energy this  energy could be a simple cosmological constant term in Einstein's equations or it could be a time dependent energy density. In any case the pressure of the dark energy must be negative to account for the acceleration.  What we would really like to know, to determine the nature of past and future acceleration, is the equation of state of the dark component, $X$, as a function of the scale parameter $R(t)$.  This equation of state is defined as 
\begin{equation}\label{eqs}
      w(z)\,\equiv\,\frac{p_X}{\rho_X},
\end{equation}
where $p_X$ is the pressure, $\rho_X$ is the energy density, and $z$ is given by the ratio of $R(t)$ to
the scale today, $R_0$, as
\begin{equation}\label{defz}
    1\,+\,z = R_0/R(t).
\end{equation}
For a cosmological constant $w(z)\,=\,-1$ for all $z$. If the acceleration is caused by extra dimensions it can also be parameterized as an effective $w(z)$.

There now exists a new data set of measurements of the luminosity distances for type Ia supernovae, which range in distance from $z\,=\,0.01$ to $z\,=\,1.755$ \cite{four}.  This data includes some new events as well as a reanalysis of the old events. There were about $194$ old events \cite{tre1,tre2} some of which have been dropped and most of the rest assigned new errors. The new set consists of $157$ ``gold'' events or $186$ gold plus ``silver'' events.

\begin{figure}[h]\centering
\includegraphics[height=3.0in]{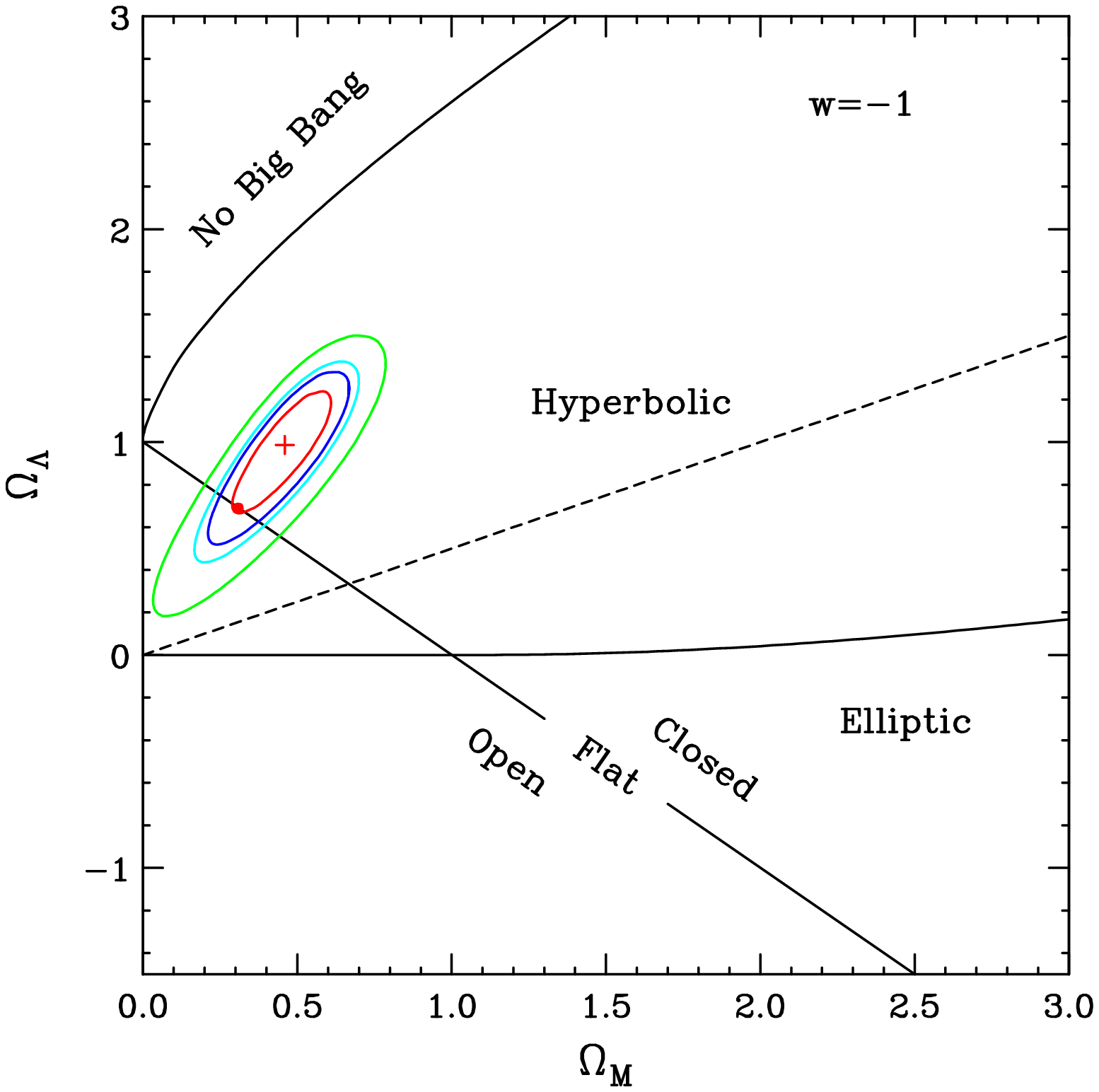}\hfill \includegraphics[height=3.0in]{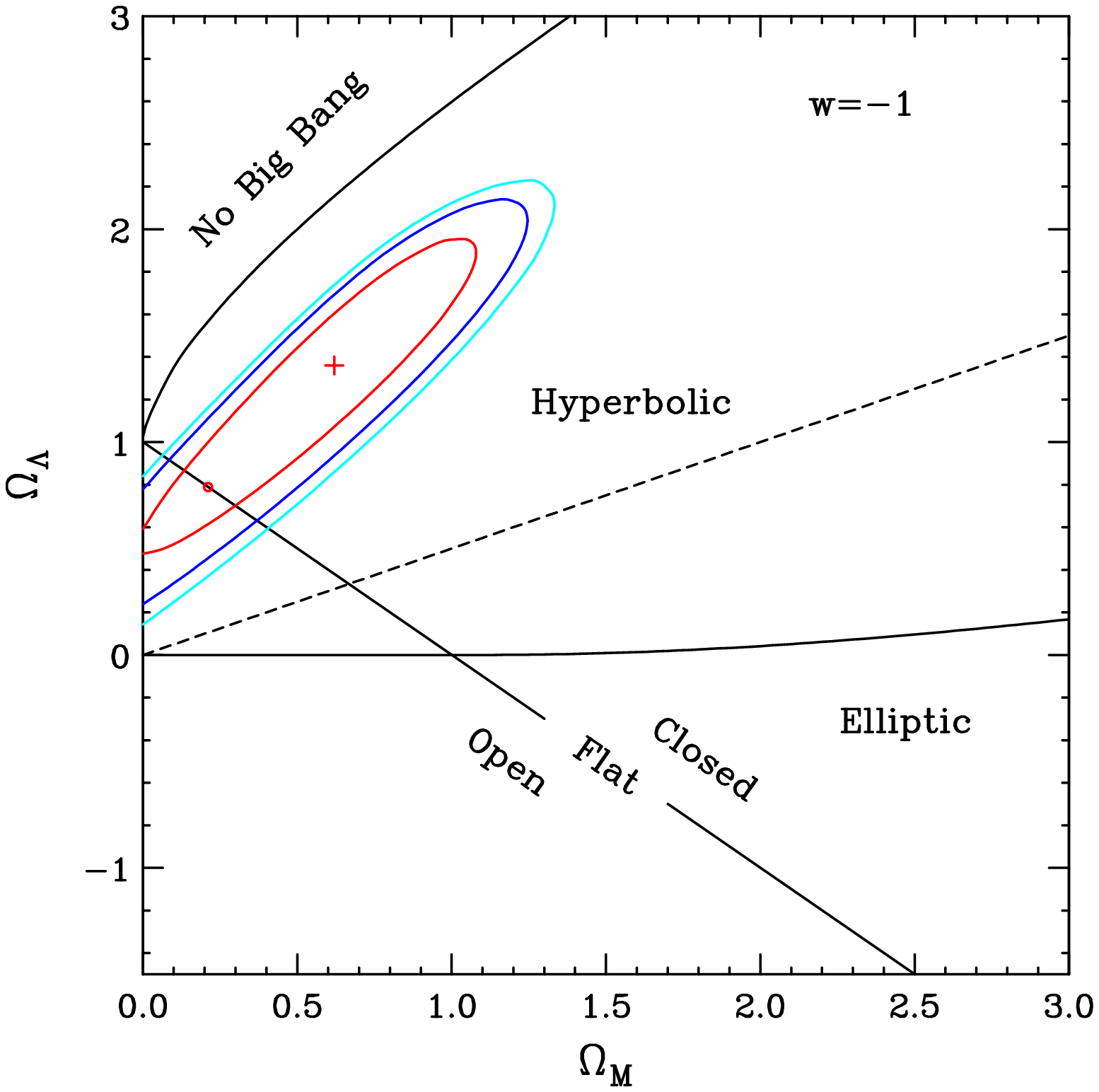}\hfill
\caption{(Color online) In the left panel, the 68.3\% (red), 90\% (dark blue), 95.4\% (blue) and 99.7\% (green) confidence contours 
for the gold data fit with a cosmological constant are shown.  The cross indicates the best fit, and the solid circle the best fit to a flat cosmology. The right panel shows the 68.3\%, 90\% and the 95.4\% confidence contours for the data in Ref.\,\cite{hiz}.} \label{omox}
\end{figure}

The gold data set gives joint confidence levels in the matter\,-\,dark energy plane with $w\,=\,-1$ that are more restrictive than the earlier data and in good agreement with other constraints \cite{four}. This is shown in Fig.\,\ref{omox}. Thus, perhaps, it is not too early to ask whether this data 
can distinguish between simple models for $w(z)$.  To investigate this question we use a few models such as $w(z)\,=\,w_0\,+\,w_1\,z$, $w(z)\,=\,w_0$ and $w(z)\,=\,w_0\,+\,w_1\,z/(1\,+\,z)$ to calculate a luminosity distance. We then minimize the $\chi^2$ of the difference between the measured and calculated luminosity distances in an attempt to determine the constant parameters $w_0$ and $w_1$. WMAP \cite{wmap} and other experiments \cite{SDSS} have accurately determined the matter density and curvature, defined as
\begin{eqnarray}
\Omega_M\,\equiv\,\frac{8\pi\,G}{3\,H^2_0}\rho_0\,,\label{Om} \\
\Omega_k\,\equiv\,-\,\frac{k}{R^2_0\,H^2_0}\,,\label{Ok} 
\end{eqnarray}
to be close to $\Omega_M\,=\,0.3$ and $\Omega_k\,=\,0$.  $\rho_0$ is the density of matter today, $H_0$ is the Hubble constant today, and $k\,=\,-1,0,+1$.  Nevertheless, because we would like to know to what extent the supernova data agrees with these other experiments, we will sometimes include $\Omega_M$ and/or $\Omega_k$ in our set of parameters to be fit despite the fact that this exacerbates the degeneracy problem that arises because the luminosity distance depends only on the sum of the matter, curvature and dark energy contributions.

In an earlier paper \cite{DR}, we explored the $\chi^2$ sensitivity of the data in Refs.\,\cite{hiz,super} to simple assumptions about the dark energy density. Recently, many other  papers have appeared \cite{fits,greek} which use the data of Refs.\,\cite{four,tre1,tre2} to analyze numerous models of the dark energy. A paper which is close in spirit to this work and which we had the advantage of reading  before we started this work is Ref.\,\cite{greek}. While we disagree with some of the results and conclusions, there is still much to like about this reference; many more models were studied and physical motivation is given for some of the models. We, on the other hand, have chosen to consider only the simplest cases to remain as model-independent as
possible.

Our results are not encouraging.  The errors in the fitted parameters are often more than $100\%$, in some cases much more. We estimate the error in $\chi^2$ itself is more than $10\%$.  Still some consistencies can be seen and some patterns discerned. And while the data is not manifestly in agreement with other experiments (e.g., $\Omega_M\,=\,0.3$ and $\Omega_k\,=\,0$\,) it is not in disagreement.  One (negative) result that seems very clear is that the absolute value of $\chi^2$ cannot be used, at this time (i.~e., from this data set), to discriminate between models for $w(z)$.

\section{Calculation of the luminosity distance and $\chi^2$}

The luminosity distance is given by
\begin{equation}\label{dL}
d_L(z)\,=\,\frac{c(1+z)}{H_0\sqrt{\ds|\Omega_k|}}\left\{\begin{array}{c}
                                                  \sin \\
                                                  1    \\
                                                  \sinh
                                                  \end{array}\right\}\left[\sqrt{|\Omega_k|}
\int_0^z\,dz'\frac{H_0}{H(z')}\right]\,,
\end{equation}
where $H(z)$ is the Hubble parameter
\begin{eqnarray}
\frac{H^2(z)}{H^2_0}\,&=&\,\Omega_M(1+z)^3\,+\,\Omega_k(1+z)^2\,+\,\Omega_X\,f(z)\,,\label{hz} \\
\Omega_X\,&=&\,1\,-\,\Omega_M\,-\,\Omega_k\,,\label{OX}
\end{eqnarray}
and $f(z)$ is given by the equation of state
\begin{equation}\label{fz}
f(z)\,=\,\exp\left\{3\,\int_0^z\frac{dx}{1\,+\,x}\left(1\,+\,w(x)\right)\right\}.
\end{equation}
In Eq.\,(\ref{dL}) $\sin[\;]$ is used for $\Omega_k\,<\,0$ ($k\,=\,+1$), $\sinh[\;]$ is used for $\Omega_k\,>\,0$ ($k\,=\,-1$), and the unmodified square bracket is used for $\Omega_k\,=\,0$.

Many references give models where $f(z)$ is not obviously related to an equation of state. However, from Eq.\,(\ref{fz}) we see that $f(0)\,=\,1$, $f(z)\,\ge\,0$, and, given an explicit expression for $f(z)$,
\begin{equation}\label{weffz}
w(z)\,=\,-1\,+\,\frac{1}{3}\left(1\,+z\,\right)\frac{f'(z)}{f(z)},
\end{equation}
where the prime denotes a derivative with respect to $z$. This equation can also be used to express the effects of extra dimensions as an effective $w(z)$.

Below we use the dimensionless luminosity distance
\begin{equation}\label{DL}
D_L(z)\,=\,\frac{H_0\,d_L(z)}{c}
\end{equation}
and the quantity that is compared to the data is $5\,\log_{10}D_L(z)\,+\,M$ where $M$ is a constant offset between the data and the theoretical expression.  The comparison is made by calculating $\chi^2$ where
\begin{equation}\label{chisq}
\chi^2\,=\,\sum_{i=1}^{N}\frac{\left[5\,\log_{10}D_L^{\mathrm{exp}}(z_i)\,-\,M\,-\,5\, \log_{10}D_L(z_i)\right]^2}{\sigma_i^2}.
\end{equation}
The sum is over the $N\,=\,157$ events contained in the gold set \cite{four} and $\sigma_i$ are the experimental errors in $5\log_{10}D_L^{\mathrm{exp}}(z_i)$.

$D_L(z)$, through the equation of state $w(z)$, Eq.\,(\ref{eqs}), depends on some set of parameters, $a_1, a_2, \ldots, a_n$, and we minimize $\chi^2$ by varying these parameters.  $\chi^2$ also depends on $M$ but to minimize with respect to $M$ is trivial,
\begin{equation}\label{M}
M\,=\,\frac{N}{D}\,,
\end{equation}
where $N$ and $D$ are given by
\begin{eqnarray}
N\,&=&\,\sum_{i=1}^N\,\frac{5\,\log_{10}D_L^{\mathrm{exp}}(z_i)\,-\,5\,\log_{10}D_L(z_i)}{\sigma_i^2}\,, \\
D\,&=&\,\sum_{i=1}^N\,\frac{1}{\sigma_i^2}.
\end{eqnarray}
Sometimes we use Eq.\,(\ref{M}) for $M$, sometimes we include $M$ as one of the parameters $a_i$ to be fit, and we have checked that we get the same result.

The data indicates that the universe was decelerating in the past and it is of interest to know when that changed.  The acceleration is given by
\begin{eqnarray}
q\,&\equiv&\,-\frac{1}{H^2R}\frac{d^2R}{dt^2}\,,\label{q1} \\
&=&\,\frac{1}{2}\frac{d\,\ln H^2}{d\,\ln(1+z)}\,-\,1\,,\label{q2}
\end{eqnarray}
and, given $H^2$ from Eq.\,(\ref{hz}), and values for the parameters, it is straightforward to find $z_*$ where $q(z_*)\,=\,0$.

\section{Results}

We consider three simple, rather ad hoc, expressions for the equation of state \cite{modelI}
\begin{eqnarray}
w(z)\,&=&\,w_0\,+\,w_1\frac{z}{1+z}\,,\label{model1} \\
w(z)\,&=&\,w_0\,,\label{model2} \\ [5pt]
w(z)\,&=&\,w_0\,+\,w_1 z\,,\label{model3}
\end{eqnarray}
where $w_0$ and $w_1$ are constants. These expressions allow us to evaluate \begin{table}[h]
\begin{center}
\begin{tabular}{| r | c | c | c | c | c | c |}
\hline
\multicolumn{1}{ | r |}{\hh \Large Model } &
\multicolumn{1}{ c |}{ \Large $\chi^2$  } &
\multicolumn{1}{ c |}{ \Large $w_0$ } &
\multicolumn{1}{ c |}{ \Large $w_1$ } &
\multicolumn{1}{ c | }{ \Large $\Omega_M$ } &
\multicolumn{1}{ c |}{ \Large $\Omega_X$ }  &
\multicolumn{1}{ c |}{ \Large $z_*$ }  \\  \hline \hline
{ } & {  } & { } & { } & { } & { } & { } \\
\raisebox{1.5ex}[0pt]{ I. } & \raisebox{1.5ex}[0pt]{ $172.3$ } & \raisebox{1.5ex}[0pt]{ $-10.3^{+7.5}_{-13.7}$ } 
& \raisebox{1.5ex}[0pt]{ $30.5^{+43.5}_{-300+}$ }
& \raisebox{1.5ex}[0pt]{ $0.27^{+0.20}_{-0.12}$ } & \raisebox{1.5ex}[0pt]{ $0.32^{+0.19}_{-0.07}$ } 
& \raisebox{1.5ex}[0pt]{ $0.14^{+0.35}_{-0.11}$ }\\ \hline
{ } & {  } & {  } & {   } & {   } & {  }  & { } \\ 
\raisebox{1.5ex}[0pt]{ II. } & \raisebox{1.5ex}[0pt]{ $173.5$  } & \raisebox{1.5ex}[0pt]{ $-2.48^{+1.50}_{-1.95}$  } 
& \raisebox{1.5ex}[0pt]{ $3.8^{+6.0}_{-16.8}$ } 
& \raisebox{1.5ex}[0pt]{ $0.46^{+0.09}_{-0.43}$ } & \raisebox{1.5ex}[0pt]{ $1\,-\,\Omega_M$ } 
& \raisebox{1.5ex}[0pt]{ $0.29^{+3.05}_{-0.26}$ } \\ \hline
{ } & {  } & {   } & {   } & {  } & {  } & { } \\ 
\raisebox{1.5ex}[0pt]{ III. } &\raisebox{1.5ex}[0pt]{ $173.9$ } & \raisebox{1.5ex}[0pt]{ $-1.58^{+0.33}_{-0.32}$ } 
& \raisebox{1.5ex}[0pt]{ $3.3^{+1.6}_{-1.7}$  } 
& \raisebox{1.5ex}[0pt]{ $0.3$  } & \raisebox{1.5ex}[0pt]{ $0.7$ } 
& \raisebox{1.5ex}[0pt]{ $0.35^{+0.17}_{-0.18}$ } \\ \hline 
{ } & { } & { } & { } & { } & { } & { } \\
\raisebox{1.5ex}[0pt]{ IV. } & \raisebox{1.5ex}[0pt]{ 172.5 } & \raisebox{1.5ex}[0pt]{ $-7.2^{+4.8}_{-9.5}$ } 
& \raisebox{1.5ex}[0pt]{ 0 } & \raisebox{1.5ex}[0pt]{ $0.33^{+0.16}_{-0.12}$ } 
& \raisebox{1.5ex}[0pt]{ $0.34^{+0.15}_{-0.07}$ }  
& \raisebox{1.5ex}[0pt]{ $0.16^{+0.29}_{-0.09}$ } \\ \hline
{ } & {  } & {  } & {  } & {   } & {  }  & { } \\ 
\raisebox{1.5ex}[0pt]{ V. } & \raisebox{1.5ex}[0pt]{ $173.7$ } & \raisebox{1.5ex}[0pt]{ $-2.40^{+0.92}_{-1.76}$  } 
& \raisebox{1.5ex}[0pt]{ $0$ } 
& \raisebox{1.5ex}[0pt]{ $0.49^{+0.05}_{-0.08}$ } & \raisebox{1.5ex}[0pt]{ $1\,-\,\Omega_M$ } 
& \raisebox{1.5ex}[0pt]{ $0.30^{+0.14}_{-0.09}$ } \\  \hline
{ } & { } & { } & { } & { } & { }  & { } \\
\raisebox{1.5ex}[0pt]{ VI.} & \raisebox{1.5ex}[0pt]{ 177.1 } & \raisebox{1.5ex}[0pt]{ $-1.02\,\pm\,0.12$ }   
& \raisebox{1.5ex}[0pt]{ 0 } 
& \raisebox{1.5ex}[0pt]{ 0.3 } & \raisebox{1.5ex}[0pt]{ 0.7 } 
& \raisebox{1.5ex}[0pt]{ $0.67\,\pm\,0.01$ } \\ \hline

{ } & { } & { } & { } & { } & { } & { } \\
\raisebox{1.5ex}[0pt]{ VII. } & \raisebox{1.5ex}[0pt]{  $172.1$ } 
& \raisebox{1.5ex}[0pt]{ $-10.0^{+6.8}_{-12.0}$ } & \raisebox{1.5ex}[0pt]{ $16^{+22}_{-225+}$ } 
& \raisebox{1.5ex}[0pt]{ $0.27^{+0.17}_{-0.13}$ } & \raisebox{1.5ex}[0pt]{ $0.32^{+0.12}_{-0.08}$ } 
& \raisebox{1.5ex}[0pt]{ $0.14^{+0.41}_{-0.10}$ } \\ \hline
{ } & {  } & {  } & {  } & {  } & { } & { } \\
\raisebox{1.5ex}[0pt]{ VIII. } & \raisebox{1.5ex}[0pt]{ $173.6$ } & \raisebox{1.5ex}[0pt]{ $-2.44^{+1.23}_{-1.91}$ } 
& \raisebox{1.5ex}[0pt]{ $1.7^{+3.8}_{-14.5}$ } 
& \raisebox{1.5ex}[0pt]{ $0.48^{+0.07}_{-0.23}$ } & \raisebox{1.5ex}[0pt]{ $1\,-\,\Omega_M$ } 
& \raisebox{1.5ex}[0pt]{ $0.29^{+0.53}_{-0.22}$ }  \\ \hline 
{ } & {  } & {  } & {  } & { } & { } & { } \\
\raisebox{1.5ex}[0pt]{ IX.} & \raisebox{1.5ex}[0pt]{ $174.4$ } & \raisebox{1.5ex}[0pt]{ $-1.40\,\pm\,0.25$ } 
& \raisebox{1.5ex}[0pt]{ $1.67^{+0.86}_{-0.95}$ } 
& \raisebox{1.5ex}[0pt]{ 0.3 } & \raisebox{1.5ex}[0pt]{ $0.7$ } 
& \raisebox{1.5ex}[0pt]{ $0.39^{+0.15}_{-0.16}$ } \\ \hline
{ } & { } & {   } & {   } & {   } & {  } & { } \\
\raisebox{1.5ex}[0pt]{ X. } & \raisebox{1.5ex}[0pt]{ $177.1$ } & \raisebox{1.5ex}[0pt]{ $-1$  } 
& \raisebox{1.5ex}[0pt]{ 0  } & \raisebox{1.5ex}[0pt]{ $0.31\,\pm\,0.04$  } & \raisebox{1.5ex}[0pt]{ $1\,-\,\Omega_M$ } 
& \raisebox{1.5ex}[0pt]{ $0.65\,\pm\,0.10$ } \\  \hline
{ } & {  } & { } & { } & { } & {  } & { }  \\
\raisebox{1.5ex}[0pt]{ XI. } & \raisebox{1.5ex}[0pt]{ $ 177.1$ } & \raisebox{1.5ex}[0pt]{ $-1$ } & \raisebox{1.5ex}[0pt]{ 0 }
& \raisebox{1.5ex}[0pt]{ 0.3 } & \raisebox{1.5ex}[0pt]{ $0.72\,\pm\,0.10$ } 
& \raisebox{1.5ex}[0pt]{ $0.69\,\pm\,0.08$ } \\ \hline
{ } & {  } & { } & { } & {  } & {  }  & { } \\
\raisebox{1.5ex}[0pt]{ XII. } & \raisebox{1.5ex}[0pt]{ $175.0$ } & \raisebox{1.5ex}[0pt]{ $-1$ } & \raisebox{1.5ex}[0pt]{ 0 } 
& \raisebox{1.5ex}[0pt]{ $0.46^{+0.10}_{-0.11}$ } & \raisebox{1.5ex}[0pt]{ $0.98^{+0.18}_{-0.20}$ }  
& \raisebox{1.5ex}[0pt]{ $0.63^{+0.25}_{-0.22}$ } \\ \hline 
\hline
\end{tabular}
\end{center}
\caption{ Fit of supernova Ia data to $w = w_0 + w_1 - w_1 y$, $y=(1+z)^{-1}$ (first 3 rows), $w=w_0$ (rows IV, V and VI), $w=w_0+w_1 z$ (rows VII, VIII, and IX) and some special cases (last three rows). The errors are the $68.3\%$ confidence level numbers. The boxes without errors mean that parameter was held fixed. If a parameter was varied no restrictions were placed on its possible values except for $\Omega_M \ge 0$ and $\Omega_X \ge 0$. The last column uses the values of the parameters and their errors to find the $z$ value where the universe changes from decelerating to accelerating.}
\label{sn1a04_1}
\end{table}
the integral in Eq.\,(\ref{fz}) and avoid having an integral within an integral. We minimize $\chi^2$ with respect to $w_0$, $w_1$ (if relevant), $\Omega_M$, $\Omega_k$ (or $\Omega_X$), or some subset of these; sometimes we hold $\Omega_k$ fixed at zero or $\Omega_M$ fixed at $0.3$ or $w_0$ fixed at $-1$ or some combination of these conditions. If a parameter is allowed to vary we impose no prior conditions on the values it may assume except for $\Omega_M\,\ge\,0$ and $\Omega_X\,\ge\,0$.

It is {\em not} straightforward to minimize $\chi^2$ because the minimum is very shallow and very broad.  We use two independent numerical codes to cross check, one is a slight generalization of programs (MRQMIN) given in Numerical Recipes \cite{NR}, the other is a code called STEPIT \cite{jpc} which is available on the web.

\begin{figure}[h]\centering
\includegraphics[height=3.0in]{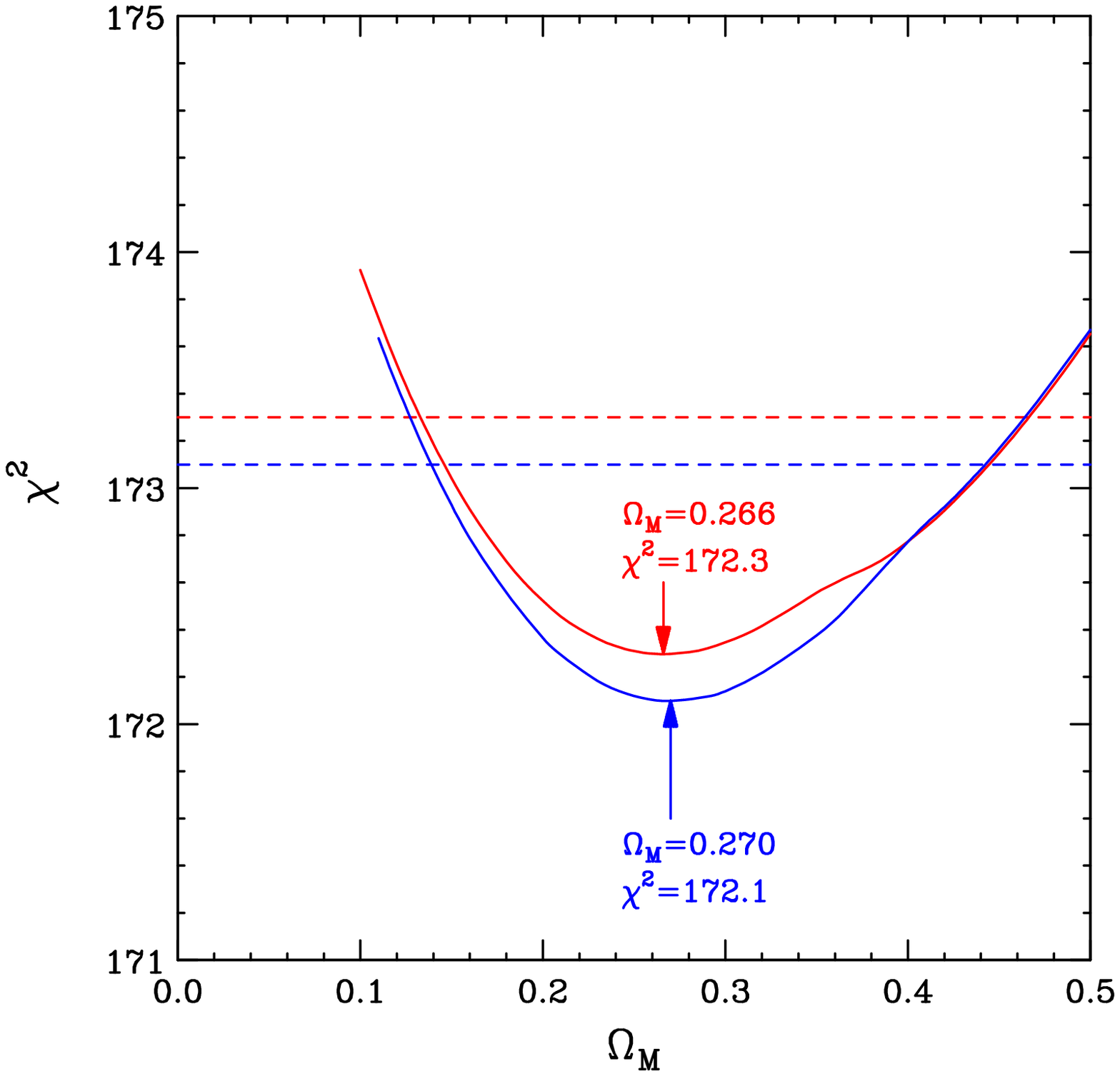}
\hfill\includegraphics[height=3.0in]{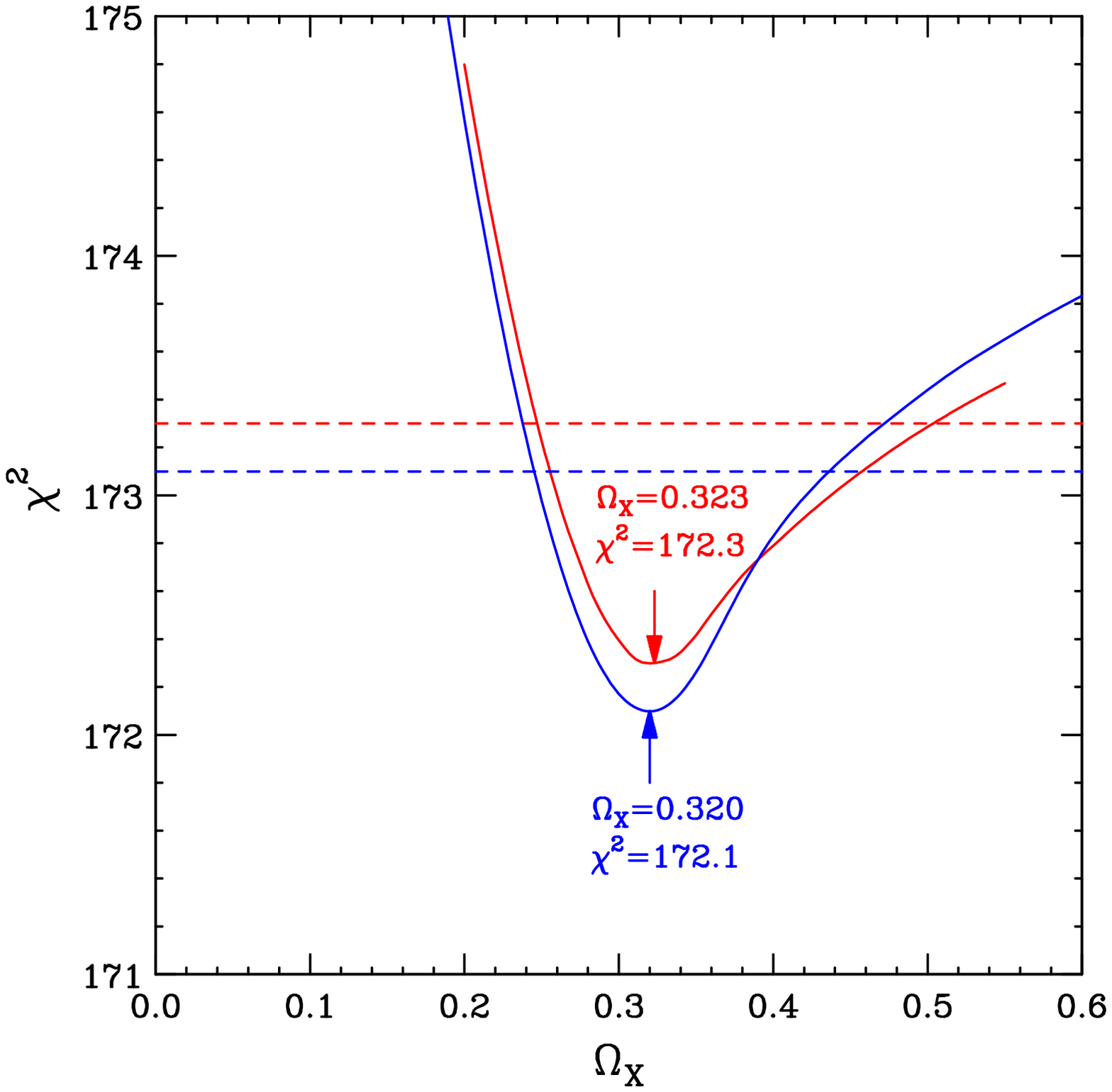}\hfill  \\
\includegraphics[height=3.0in]{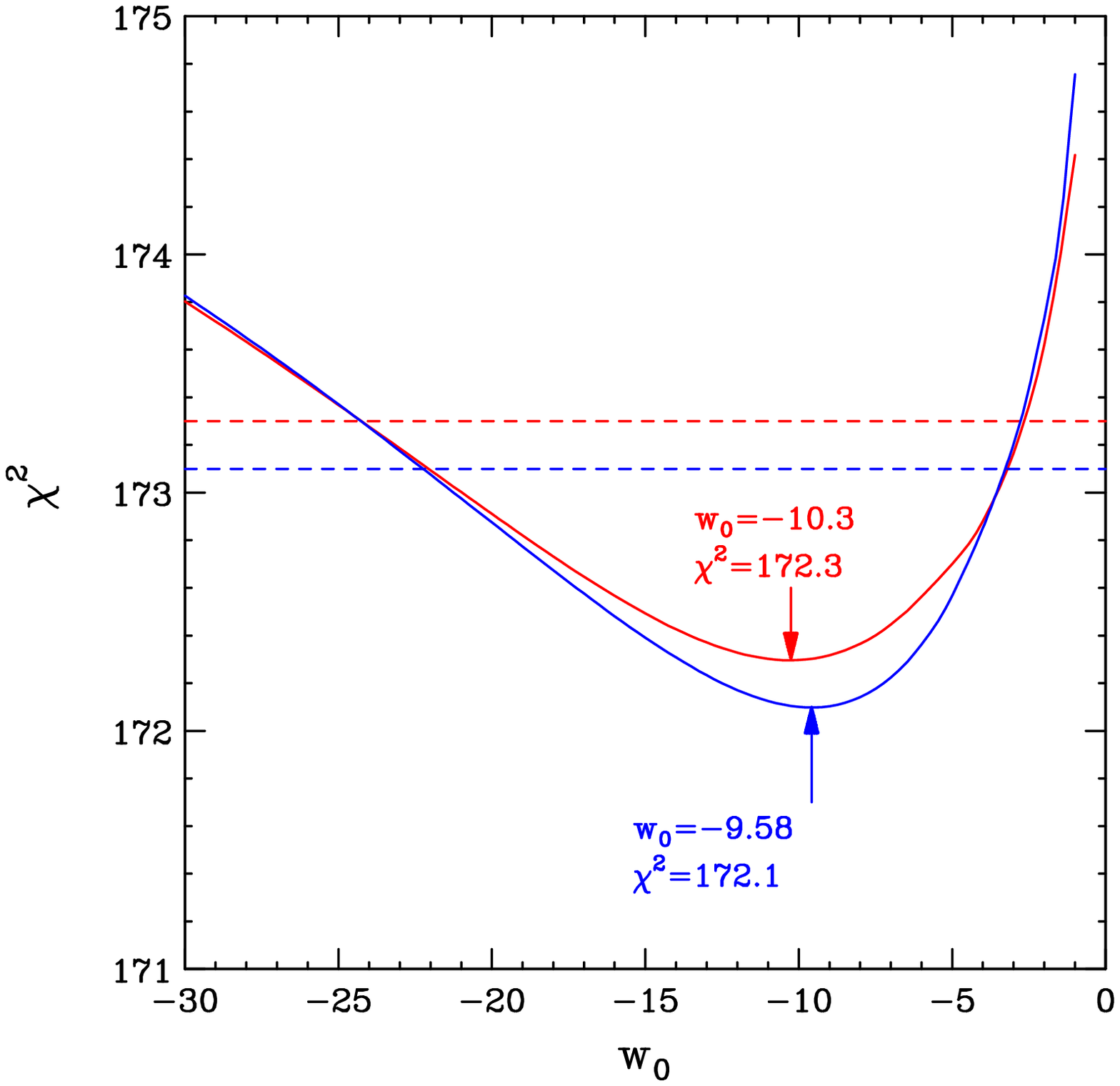}
\hfill\includegraphics[height=3.0in]{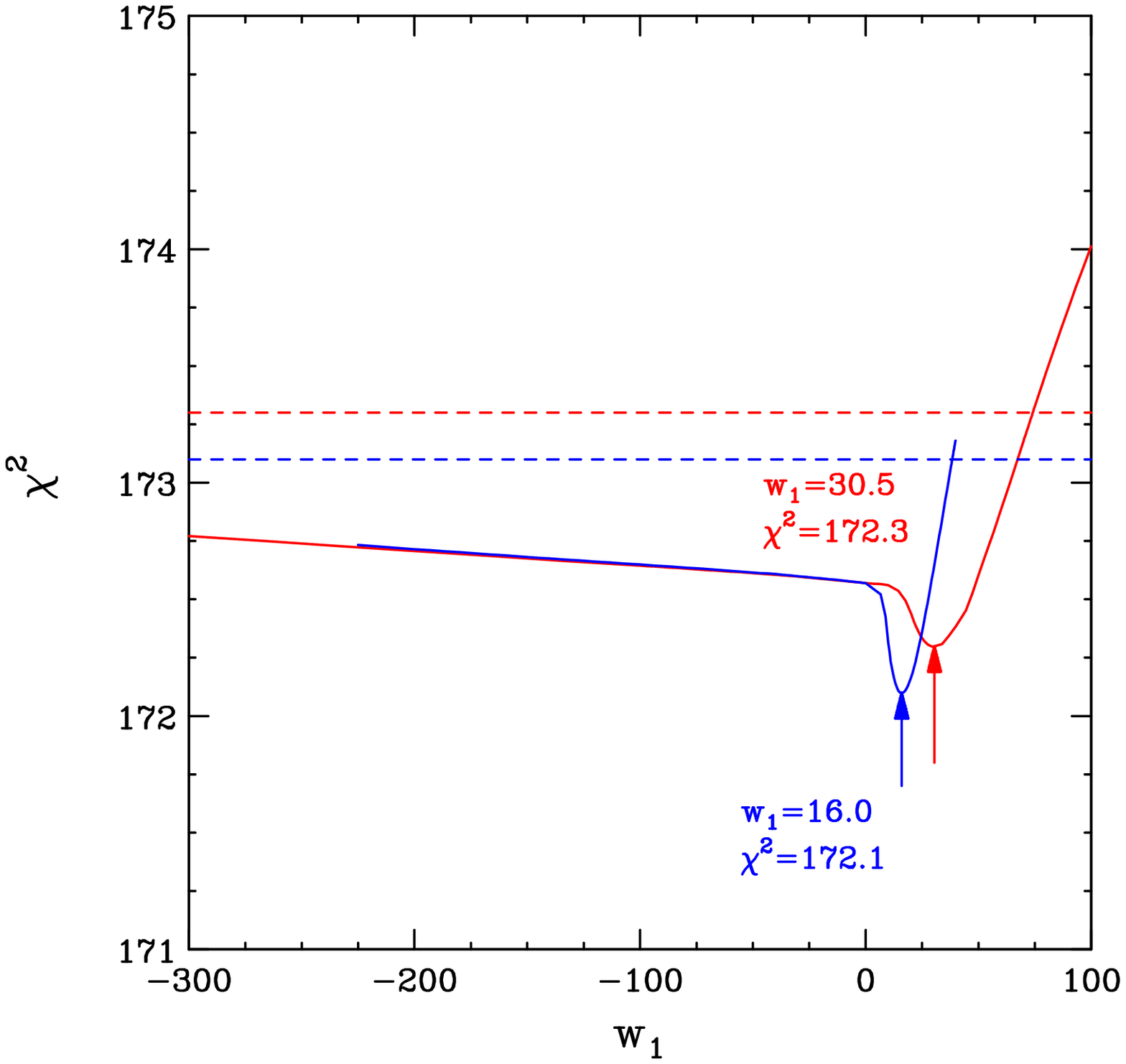}\hfill
\caption{(Color online) The $\chi^2$ variation of the parameters $\Omega_M$, $\Omega_X$, $w_0$ and $w_1$ is shown for the cases
$w(z)\,=\,w_0\,+\,w_1\,z/(1\,+\,z)$ (red) and $w(z)\,=\,w_0\,+\,w_1\,z$ (blue).  The horizonal dashed lines
correspond to $\Delta\chi^2\,=\,1$.}\label{OmOxw0w1}
\end{figure}

\begin{figure}[h]\centering
\includegraphics[height=3.0in]{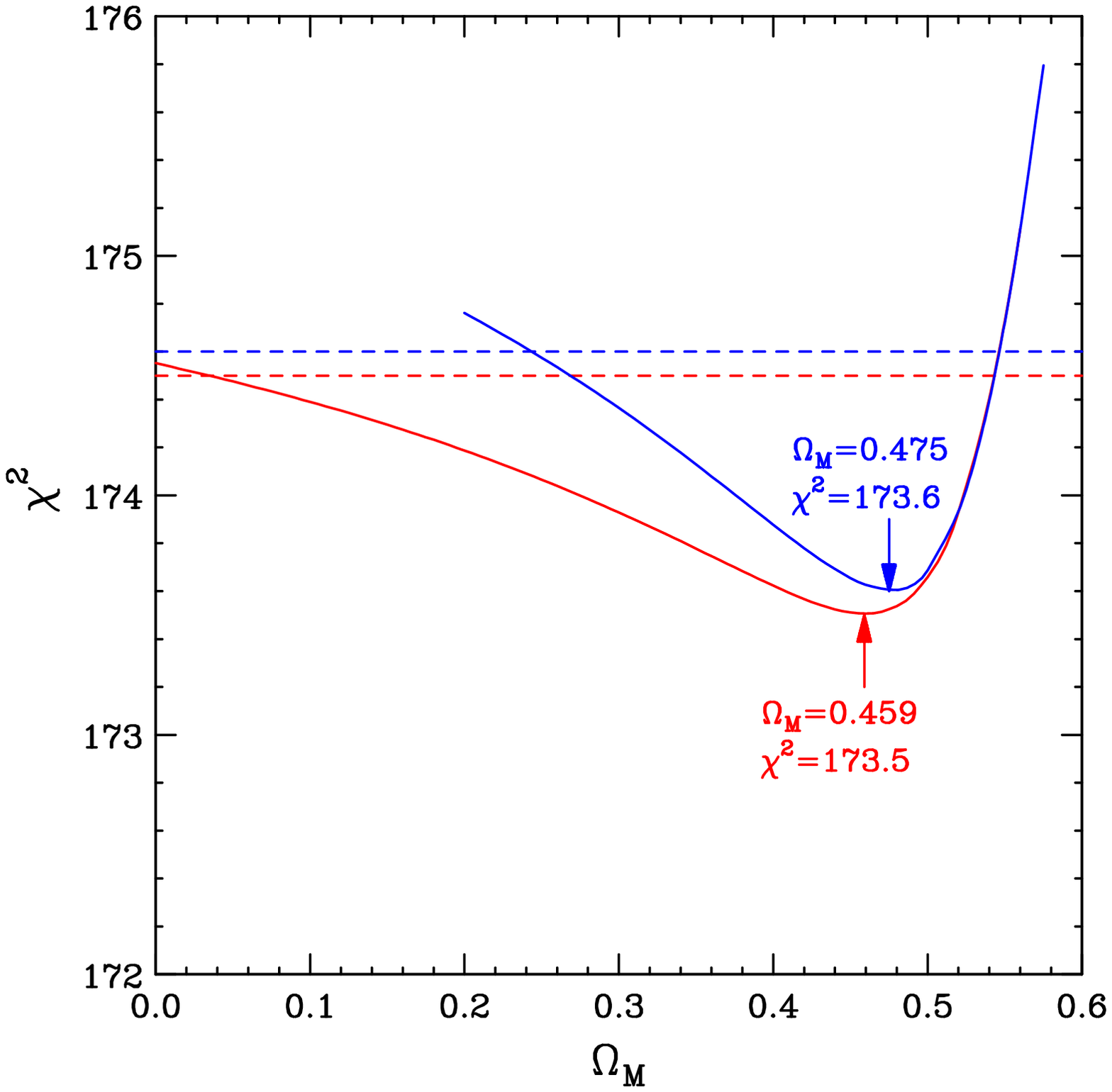}
\hfill\includegraphics[height=3.0in]{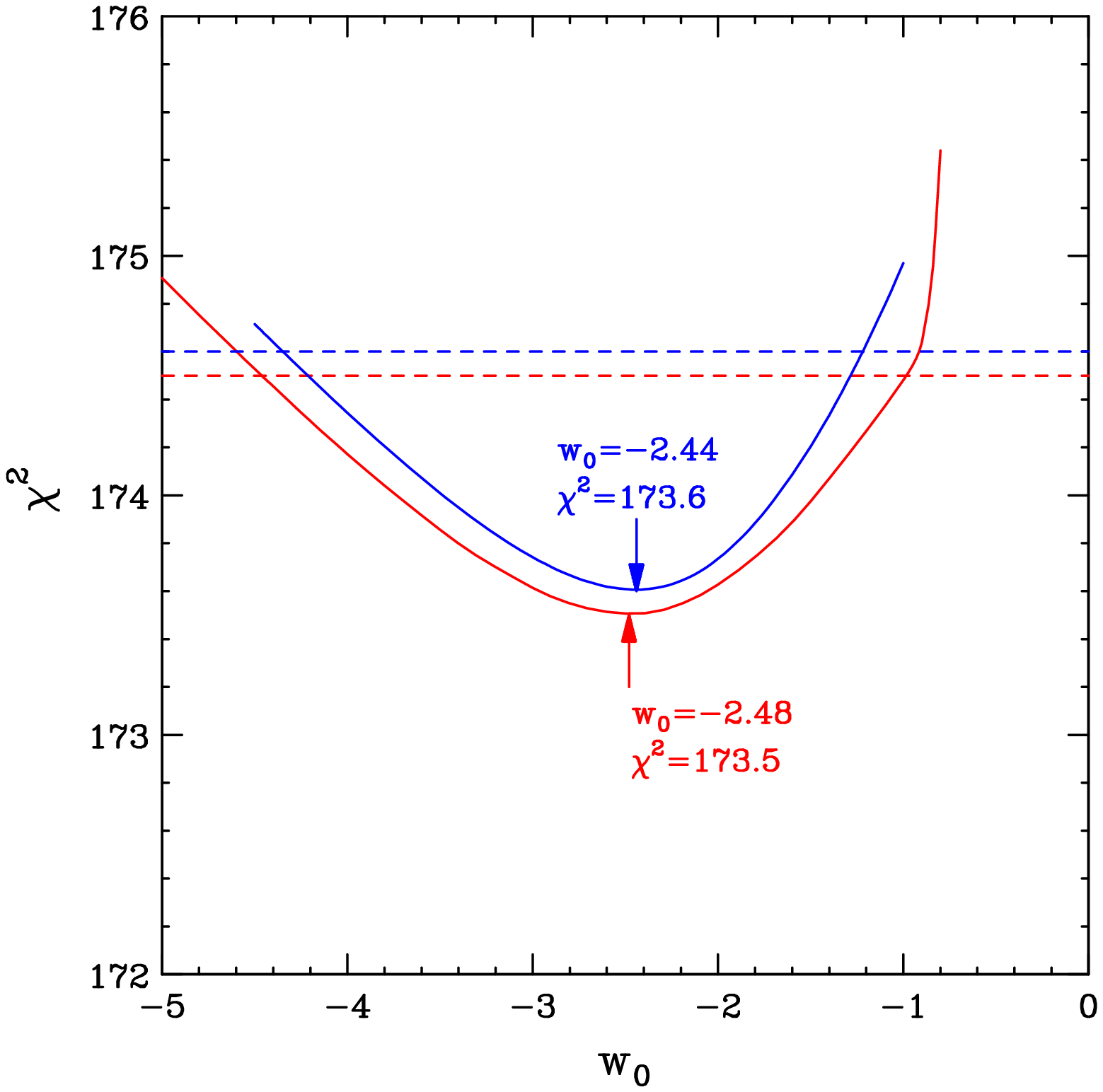}\hfill 
\hfill\includegraphics[height=3.0in]{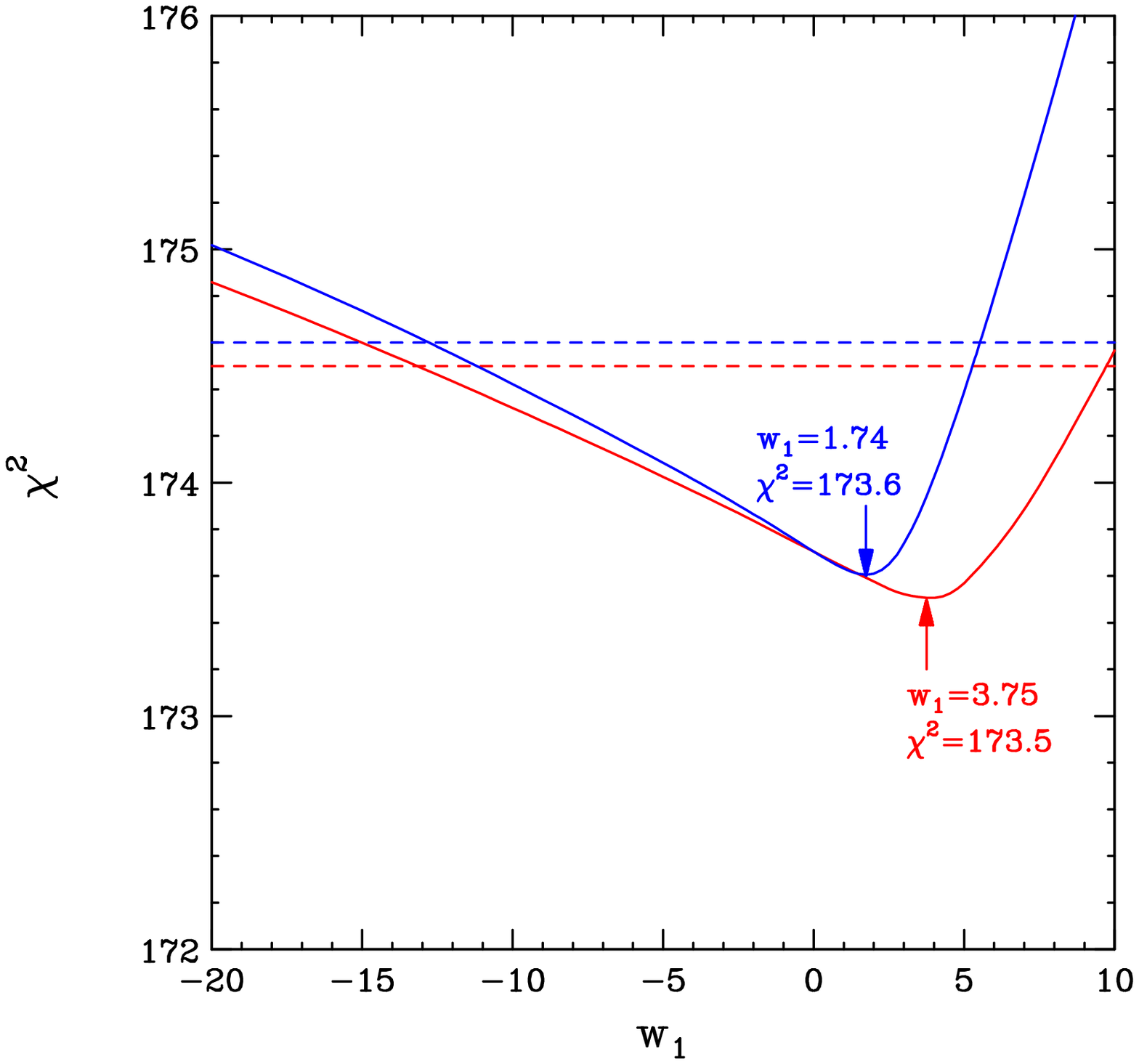}\hfill
\caption{(Color online) Same as Fig.\ref{OmOxw0w1} except that $\Omega_X\,=\,1\,-\,\Omega_M$.}\label{Omw0w1}
\end{figure}

Our results are given in Table\,\ref{sn1a04_1}.  The first 3 rows use Eq.\,(\ref{model1}), with either nothing, $\Omega_k$, or $\Omega_M$ and $\Omega_k$ fixed.  The next 6 rows do the same for Eq.\,(\ref{model2}) and then for Eq.\,(\ref{model3}). The last 3 rows give some special cases.  For each model we give the minimum $\chi^2$ value and the values of the parameters which give that value.  The errors on the parameters are the $68.3\%$ confidence level uncertainties given for a particular parameter by finding the variation in that parameter which results in an increase of $\chi^2_{Min}$ by $1$ when the other parameters are allowed to vary freely.  Plots of how $\chi^2$ varies to the $1\,\sigma$ error in each parameter are shown in Fig.\,\ref{OmOxw0w1} for Models I (red curves)and VII (blue curves). If the 
parameters in these models are restricted to the case of a flat cosmology, $\Omega_k\,=\,0$, the resulting $1\,\sigma$ errors of Models II and VIII are shown in Fig.\,\ref{Omw0w1}.

To try to get an estimate of the uncertainty in $\chi^2_{Min}$ itself, and to check the stability of the numbers in Table\,\ref{sn1a04_1}, we have done approximately $120$ bootstrap simulations.  These simulations consist of constructing new data sets by randomly selecting $157$ events from the original data set.  We do enough such simulations  that each event gets used about the same number of times. We then calculate the average $\chi^2_{Min}$ for the simulations and, to find an error for $\chi^2_{Min}$, take the standard deviation of the set of $\chi^2_{Min}$.  We also find the average and the standard deviation for each of the parameters. These results are given in Table\,\ref{sn1a04_2}. The values of the parameters and 
\begin{table}[h]
\begin{center}
\begin{tabular}{| r | c | c | c | c | c | }
\hline
\multicolumn{1}{ | r |}{\hh \Large Model } &
\multicolumn{1}{ c |}{ \Large $\chi^2$  } &
\multicolumn{1}{ c |}{ \Large $w_0$ } &
\multicolumn{1}{ c |}{ \Large $w_1$ } &
\multicolumn{1}{ c | }{ \Large $\Omega_M$ } &
\multicolumn{1}{ c |}{ \Large $\Omega_X$ }  \\  \hline \hline
{ } & { } & { } & { } & { } & { } \\
\raisebox{1.5ex}[0pt]{ Ib. } & \raisebox{1.5ex}[0pt]{ $169.8\,\pm\,19.6$ } & \raisebox{1.5ex}[0pt]{ $-12.5\,\pm\,14.2$ }
& \raisebox{1.5ex}[0pt]{ $3.8\,\pm\,122.5$ }
& \raisebox{1.5ex}[0pt]{ $0.28\,\pm\,0.18$ } & \raisebox{1.5ex}[0pt]{ $0.36\,\pm\,0.11$ } \\ \hline
{ } & { } & { } & { } & { } & { } \\
\raisebox{1.5ex}[0pt]{ IIb. } & \raisebox{1.5ex}[0pt]{ $171.1\,\pm\,20.3$ } & \raisebox{1.5ex}[0pt]{ $-3.05\,\pm\,3.38$ }
& \raisebox{1.5ex}[0pt]{ $0.65\,\pm\,20.7$ }
& \raisebox{1.5ex}[0pt]{ $0.42\,\pm\,0.15$ } & \raisebox{1.5ex}[0pt]{ $1\,-\,\Omega_M$ } \\ \hline
{ } & { } & { } & { } & { } & { } \\
\raisebox{1.5ex}[0pt]{ IIIb. } & \raisebox{1.5ex}[0pt]{ $173.2\,\pm\,19.6$ } & \raisebox{1.5ex}[0pt]{ $-1.57\,\pm\,0.34$ }
& \raisebox{1.5ex}[0pt]{ $3.2\,\pm\,1.8$ }
& \raisebox{1.5ex}[0pt]{ $0.3$ } & \raisebox{1.5ex}[0pt]{ $0.7$ } \\ \hline
{ } & {  } & { } & { } & { } & { } \\
\raisebox{1.5ex}[0pt]{ IVb. } & \raisebox{1.5ex}[0pt]{ $170.0\,\pm\,20.2$ } & \raisebox{1.5ex}[0pt]{ $-8.3\,\pm\,7.7$ } 
& \raisebox{1.5ex}[0pt]{ $ 0 $ }
& \raisebox{1.5ex}[0pt]{ $0.34\,\pm\,0.16$ } & \raisebox{1.5ex}[0pt]{ $0.55\,\pm\,0.41$ } \\ \hline
{ } & { } & { } & { } & { } & { } \\
\raisebox{1.5ex}[0pt]{ Vb. } & \raisebox{1.5ex}[0pt]{ $171.3\,\pm\,20.2$ } & \raisebox{1.5ex}[0pt]{ $-3.30\,\pm\,3.19$ }
& \raisebox{1.5ex}[0pt]{ $ 0 $ }
& \raisebox{1.5ex}[0pt]{ $0.47\,\pm\,0.10$ } & \raisebox{1.5ex}[0pt]{ $1\,-\,\Omega_M$ } \\ \hline
{ } & { } & { } & { } & { } & { } \\
\raisebox{1.5ex}[0pt]{ VIb. } & \raisebox{1.5ex}[0pt]{ $177.2\,\pm\,19.5$ } & \raisebox{1.5ex}[0pt]{ $-1.02\,\pm\,0.12$ }
& \raisebox{1.5ex}[0pt]{ $ 0 $ } & \raisebox{1.5ex}[0pt]{ $0.3$ } & \raisebox{1.5ex}[0pt]{ $0.7$ } \\ \hline
{ } & {  } & {  } & {   } & {   } & {  }  \\ 
\raisebox{1.5ex}[0pt]{ VIIb. } & \raisebox{1.5ex}[0pt]{ $170.0\,\pm\,19.4$  } & \raisebox{1.5ex}[0pt]{ $-11.2\,\pm\,10.5$  } 
& \raisebox{1.5ex}[0pt]{ $-1.4\,\pm\,82.8$ } 
& \raisebox{1.5ex}[0pt]{ $0.29\,\pm\,0.18$ } & \raisebox{1.5ex}[0pt]{ $0.39\,\pm\,0.20$ } \\ \hline
{ } & { } & { } & { } & { } & { } \\
\raisebox{1.5ex}[0pt]{ VIIIb. } & \raisebox{1.5ex}[0pt]{ $172.4\,\pm\,19.6$ } & \raisebox{1.5ex}[0pt]{ $-3.10\,\pm\,3.22$ }
& \raisebox{1.5ex}[0pt]{ $1.4\,\pm\,6.3$ }
& \raisebox{1.5ex}[0pt]{ $0.43\,\pm\,0.14$ } & \raisebox{1.5ex}[0pt]{ $1\,-\,\Omega_M$ } \\ \hline
{ } & { } & { } & { } & { } & { } \\
\raisebox{1.5ex}[0pt]{ IXb. } & \raisebox{1.5ex}[0pt]{ $173.7\,\pm\,19.6$ } & \raisebox{1.5ex}[0pt]{ $-1.40\,\pm\,0.26$ }
& \raisebox{1.5ex}[0pt]{ $1.64\,\pm\,0.95$ } & \raisebox{1.5ex}[0pt]{ $ 0.3 $ } & \raisebox{1.5ex}[0pt]{ $0.7$ } \\ \hline
{ } & { } & { } & { } & { } & { } \\
\raisebox{1.5ex}[0pt]{ Xb. } & \raisebox{1.5ex}[0pt]{ $177.2\,\pm\,19.4$ } & \raisebox{1.5ex}[0pt]{ $-1$ }
& \raisebox{1.5ex}[0pt]{ $0$ } & \raisebox{1.5ex}[0pt]{ $0.31\,\pm\,0.04$ } & \raisebox{1.5ex}[0pt]{ $1\,-\,\Omega_M$ } \\ \hline
{ } & { } & { } & { } & { } & { } \\
\raisebox{1.5ex}[0pt]{ XIb. } & \raisebox{1.5ex}[0pt]{ $177.2\,\pm\,19.4$ }& \raisebox{1.5ex}[0pt]{ $-1$ } 
& \raisebox{1.5ex}[0pt]{ $0$ } & \raisebox{1.5ex}[0pt]{ $0.3$ } & \raisebox{1.5ex}[0pt]{ $0.72\,\pm\,0.10$ } \\ \hline
{ } & { } & { } & { } & { } & { } \\
\raisebox{1.5ex}[0pt]{ XIIb. } & \raisebox{1.5ex}[0pt]{ $174.2\,\pm\,19.5$ } & \raisebox{1.5ex}[0pt]{ $-1$ }
& \raisebox{1.5ex}[0pt]{ $0$ }
& \raisebox{1.5ex}[0pt]{ $0.46\,\pm\,0.11$ } & \raisebox{1.5ex}[0pt]{ $0.98\,\pm\,0.19$ } \\ \hline
\hline
\end{tabular}
\end{center}
\caption{The averages of approximately 120 bootstrap simulations for the models of Table I. The errors are the standard deviations.}
\label{sn1a04_2}
\end{table}
the errors in the parameters are very consistent with those given in Table\,\ref{sn1a04_1}. In fact for the cases where only one or two parameters are fit they are almost identical.  The values of $\chi^2$ for the individual simulations vary from $\sim\,110$ to $\sim\,230$ with a standard deviation of about 20.  Thus, from these simulations, we conclude that the values in Table\,\ref{sn1a04_1} are robust and that $\chi^2_{Min}$ is uncertain by $10\,-\,15\%$.

\section{Conclusions}

\begin{figure}[h]\centering
\includegraphics[height=3.0in]{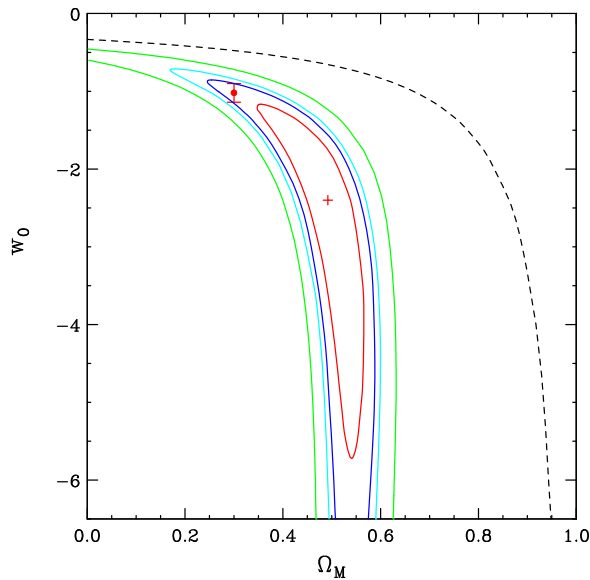}
\caption{(Color online) The 68.3\% (red), 90\% (dark blue), 95.4\% (blue) and 99.7\% (green) confidence contours for the case $\Omega_X=1-\Omega_M$ 
and $p_x=w_0\rho_x$ (model V) are shown.  The cross indicates the best fit, and the solid circle is the fit requiring $\Omega_M=0.3$. The dashed line separates accelerating (to the left) from decelerating (to the right) cosmologies.}\label{omw0}
\end{figure}

If we take the cosmological constant case ($w_0\,=\,-1$, $w_1\,=\,0$) and fit $\Omega_M$ {\em or}\, $\Omega_k$ we get the preferred answer of $0.3$ or $0$ (see Models X and XI in Table\,\ref{sn1a04_1}).  Or if we hold $\Omega_M\,=\,0.3$ and $\Omega_k\,=\,0$ and fit $w_0$ we get the preferred answer of $-1$ (Model VI). However, if we try to fit even two of the three $w_0, \Omega_M, \Omega_k$ as in Models V or XII, the fitted values are not the preferred ones although  the errors are large enough to make the parameters consistent with the desired values. This can be seen in some detail in the $w_0-\Omega_M$ confidence contours for Model V, shown in Fig.\,\ref{omw0}. Model XII, shown in more detail in Fig.\,\ref{omox},  was our original 
motivation.

Our Models III and IX are models 9 and 7 of Ref.\,\cite{greek}.  That paper used the 2003 data set \cite{tre1,tre2}; when we ran with that data set, and included the extra error due to the peculiar velocity as in \cite{greek}, we were able to find a smaller value for $\chi^2_{Min}$ in both of these cases; perhaps we used a different expression for the pecular velocity error. In any case a smaller $\chi^2_{Min}$ is not very important in view of the large uncertainty in $\chi^2_{Min}$ indicated by the simulations. What is important, in view of this uncertainty, is that small relative values of $\chi^2_{Min}$ do not indicate which models are a better fit. Also the ability to find the lowest $\chi^2$ is essential to determining the errors on the parameters; if the lowest $\chi^2$ is not found the errors will appear to be smaller than they actually are. We found much larger errors for the parameters than those given in Ref.\,\cite{greek} and this seems not to be due to uncertainity in handling the pecular velocity error.  For example our result for Model III is $w_0\,=\,-1.74\,\pm\,0.39$, $w_1\,=\,5.04\,\pm\,2.25$.

If we include structure in $w(z)$, as in Models I, II, III, which use $w(z)\,=\,w_0\,+\,w_1\,z/(1\,+\,z)$ or VII, VIII, IX, which use $w(z)\,=\,w_0\,+\,w_1\,z$, then the errors get very large and it is hard to draw any conclusions even if $\Omega_M$ and $\Omega_k$ are fixed.  What does seem clear, from the similarity of the results for Models I and VII, II and VIII, and III and IX is that low $z$ events are still dominating.  ($w_1$ of Models VII, VIII, and IX should be compared with $w_0\,+\,w_1$ of Models I, II, and III.) The factor of $(1\,+\,z)$ in the denominator of Eq.\,(\ref{model1}) is effectively just $1$. We have checked that this is also true for the gold plus silver data set. Thus comparing models for the equation of state of the dark energy will remain something of a mug's game until there exists substantially more data at higher values of $z$.

\acknowledgments We thank Rahul Malhotra, Megan Donahue, Mark Voit and Steve Zepf for useful discussions and Vinod Johri for a helpful communication. This research was supported in part by the National Science Foundation under Grant PHY-0244789 and by the United States Department of Energy under Contract No. DE-FG03-93ER40757.

\end{document}